\documentclass[prd, twocolumn, showpacs, superscriptaddress, floatfix, nofootinbib]{revtex4}

\usepackage[dvips]{graphicx}
\usepackage{epsf}
\usepackage{amsmath}
\usepackage{amssymb}

\usepackage{graphicx}
\usepackage{dcolumn}
\usepackage{bm}
\def\be{\begin{equation}}
\def\ee{\end{equation}}
\def\bea{\begin{eqnarray}}
\def\eea{\end{eqnarray}}

\usepackage{color}

\begin{document}


\title{The 21cm Signature of a Cosmic String Loop}

\author{Michael Pagano}
\email{paganom@physics.mcgill.ca}
\affiliation{Department of Physics, McGill University, Montr\'eal, QC, H3A 2T8, Canada}
\author{Robert Brandenberger}
\email{rhb@physics.mcgill.ca}
\affiliation{Department of Physics, McGill University, Montr\'eal, QC, H3A 2T8, Canada}


\begin{abstract}

Cosmic string loops lead to nonlinear baryon overdensities at early times, even before the time
which in the standard LCDM model corresponds to the time of reionization. These overdense
structures lead to signals in 21cm redshift surveys at large redshifts. In this paper, we 
calculate the amplitude and shape of the string loop-induced 21cm brightness temperature.
We find that a string loop leads to a roughly elliptical region in redshift space with
extra 21cm emission. The excess brightness temperature for strings with a tension close to 
the current upper bound can be as high as $1^{o} {\rm K}$
for string loops generated at early cosmological times (times comparable to the time
of equal matter and radiation) and observed at a redshift of $z + 1 = 30$.
The angular extent of these predicted ``bright spots'' is about $0.1^{\circ}$ for a
value of the string tension equal to the current upper bound. These signals
should be detectable in upcoming high redshift 21cm surveys.
 
\end{abstract}

\maketitle

\newcommand{\eq}[2]{\begin{equation}\label{#1}{#2}\end{equation}}

\section{Introduction}

21cm surveys promise to become an excellent window to observationally probe the
``dark ages", the epoch in the evolution of the early universe before the main burst of
star formation sets in which is the time of reionization in the standard LDCM cosmology.
The reason is that the intensity of the redshifted 21cm radiation which we receive tracks
the distribution of baryons and does not require the baryons to have formed stars.

From an experimental point of view the situation is promising. The LOFAR telescope
array has been commissioned and has the capability of providing 21cm maps between
redshift 5 and 100 at excellent angular and redshift resolution \cite{LOFAR}. LOFAR
can in turn be viewed as a prototype for the ``Square Kilometer Array" project which
will have even higher resolution \cite{SKA}, and will also be able to probe
up to redshifts of 30.

From the point of view of cosmology theory it is an interesting challenge to see
whether the current standard paradigm of early universe cosmology, the LCDM model,
will be consistent with the data. In this scenario, the primordial fluctuations
form a Gaussian random field with a scale-invariant spectrum with a slight red
tilt (more structure on larger scales). In this context, nonlinearities form
rapidly on an extended range of scales at late times, comparable and later
to the time of reionization. Observing large amplitude primordial fluctuations
at high redshifts would pose a serious problem for the standard cosmological
scenario.

In contrast, if our theory of matter is of a type which admits topological defects
such as cosmic strings, these defects will lead to nonlinearities at very high
redshifts. It was recently realized that cosmic string wakes lead to a signatures
in 21cm redshift maps with a large amplitude and with a specific geometrical
shape \cite{us}. Cosmic string wakes produced at the time $t_{eq}$ of equal
matter and radiation, with string tension close to the current upper limit
and seen at redshift $z + 1 = 30$ lead to wedges in 21cm redshift maps
with an absorption signal characterized by a brightness temperature
$\delta T_b \sim 100 {\rm mK}$. The wedges are extended in both
angular directions (a fraction of one degree) and thin in redshift direction
($\delta \nu / \nu \sim 5 \times 10^{-6}$). Such a signal should easily be
detectable to the SKA experiment.

It has been realized that many particle physics models beyond the ``Standard Model"
give rise to cosmic strings (see \cite{VilShell, HK, RHBrev} for reviews). In models
which admit stable cosmic string solutions, a network of such strings will
inevitably form during a symmetry breaking phase transition in the early universe.
The network will then approach a ``scaling solution'' in which the statistical
properties of the string network are independent of time if all lengths are scaled
to the Hubble radius $t$ (where $t$ is physical time), and will persist to
the current time. Since strings carry energy density, their gravitational
interactions will induce cosmological effects. The cosmological effects
of a string increase in amplitude as the string tension (which equals
the mass per unit length) $\mu$ increases.
The string tension, in turn, is related to the energy scale $\eta$ of the
particle physics which leads to string formation via $\mu \simeq \eta^2$.
Hence, searching for cosmological signatures of strings is a way to probe
physics beyond the ``Standard Model'' of particle physics at the high energy
scale. It is complementary to efforts to look for physics beyond the
Standard Model in terrestrial accelerators such as the LHC, experiments
which probe the low energy end of beyond the Standard Model physics.
It is possible that cosmic strings formed at a very high energy scale
are redshifted outside our current Hubble radius by a period of cosmological
inflation \cite{Guth} which occurred after string formation. However, it
has recently been realized that in fact cosmic strings are formed at the end 
of the phase of inflation in many models of inflation
based both on string theory \cite{Tye} and on more conventional supergravity 
\cite{Rachel}. Hence, there is good motivation to search for
signatures of cosmic strings even in the context of cosmological
evolution with an inflationary phase \footnote{Cosmic strings may also
arise in the form of cosmic superstrings \cite{Witten} in alternatives
to inflation such as ``String Gas Cosmology'' \cite{SGC}.} There are
already cosmological limits from the angular power spectrum of cosmic
microwave background (CMB) anisotropy experiments which set an upper
bound \cite{Dvorkin} (see also \cite{limits} for earlier
work) on the dimensionless value $G \mu$ of $\mu$ (where $G$ is Newton's
gravitational constant) which is \footnote{It is important to point
out that this limit is sensitive to parameters such as the number of
strings per Hubble volume which is currently not accurately known. An
advantage of looking for direct signatures of strings in position
space (rather than in terms of a power spectrum) is that one can obtain
bounds on $G \mu$ which are insensitive to the abovementioned paramters.
Preliminary studies of edge detection algorithms \cite{Amsel} which search for 
the signatures of cosmic strings in position space in CMB anisotropy
maps have provided evidence \cite{Danos} that the limits can be improved
by one order of magnitude.}
\be \label{constraint}
G \mu \, < \, 1.5 \times 10^{-7} \, .
\ee

The scaling distribution of cosmic strings consist of a network of infinite
strings with mean curvature radius comparable to the Hubble radius, plus
a distribution of loops with radii $R < t$. A long string segment gives
rise to a cosmic string wake whose 21cm signals were investigated in
previous work \cite{us}. In this paper, we discuss the 21cm signature of
a cosmic string loop. Cosmic string loops accrete surrounding matter
in a roughly spherical pattern \cite{early} and lead to overdensities
of matter (both dark matter and baryons) which form when the string loop
is produced at very early time and then grow in size as time progresses.
Thus, string loops will lead to large 21cm redshift signals at early
times (long before the time of reionization in the standard LCDM scenario).
The distribution of these signals is highly non-Gaussian since the
string loop distribution is not Gaussian.

The outline of this paper is as follows: In the next section we
discuss some preliminaries both concerning the distribution of cosmic
strings and concerning 21cm cosmology. Section 3 summarizes our
computation of the brightness temperature pattern induced by an 
individual cosmic string loop. We end with a discussion of our
results. 

\section{Preliminaries}

In this paper we will calculate the 21cm signature of a single
cosmic string loop. The loop distribution is determined
by the cosmic string scaling solution, according to which at
all times $t$ (sufficiently long after the phase transition
which gives rise to the strings) there will be a network
of infinite strings with mean curvature radius comparable
to the Hubble radius $t$. Since the strings are relativistic
objects, string segments will typically be moving with
relativistic velocity. Hence, there will be frequent intersections
of strings. As can be argued analytically (see \cite{VilShell, HK, RHBrev}
for reviews), the scaling solution of the infinite strings is
maintained by the process of interactions between segments of
infinite strings producing loops. 

If the infinite strings
have negligible substructure, then the loops which they produce
at time $t$ will be generated with a typical radius of 
\be
R_{form}(t) \, = \, \alpha t \, ,
\ee
where $\alpha < 1$ is a numerical constant (whose value must
be determined numerically and is at the present time still quite
uncertain). Once formed, loops will oscillate, emit gravitational
radiation and gradually shrink in size. Numerical simulations of
the dynamics of cosmic string networks \cite{CSsimuls} have
confirmed that the loop distribution approaches a scaling solution
as well, but the detailed structure of the scaling solution
is still poorly understood (see e.g. \cite{Pol} for some recent
progress). Looking for position space signatures of individual
string loops has the advantage over approaches based on computing
correlation functions that the analysis is much less dependent
on the unknown parameters entering into the description of the
string loop scaling solution.

We will be working in terms of the ``one scale model'' for the
string loop distribution (see e.g. \cite{BT}) which is based
on assuming that all string loops are born with the same radius
and that their number density redshifts as the universe expands
with the radius very slowly decreasing due to gravitational
radiation. The result for the number density $n(R, t)$ of
strings per unit volume for unit $R$ is hence
\bea
n(R, t) \, &=& \, \nu R^{-2} t^{-2} \,\,\,\,\,  R > \alpha t_{eq} \nonumber \\
n(R, t) \, &=& \, \nu R^{-5/2} t^{-2} t_{eq}^{1/2} \,\,\,\, 
\gamma G \mu t < R < \alpha t_{eq} \\
n(R, t) \, &=& \, {\rm const} \,\,\,\,\,\,\,  R < \gamma G \mu \, , \nonumber
\eea
where $\gamma$ is a constant determined by the strength of 
gravitational radiation from a string loop and which is
in the range $\gamma \sim 10$ (with substantial uncertainty).
The cosmic string loops which dominate the mass in strings (and
hence which dominate the baryonic mass accreted onto string
loops) are hence those with $R \sim \gamma G \mu$.

Since cosmic strings carry energy, they will have gravitational
effects which in turn lead to cosmological signals. Most of the
signatures of infinite string segments originate from the fact
that space perpendicular to a long straight string segment is
conical, i.e. like ${\cal{R}}^2$ with a missing wedge, the ``missing
angle'' being $8 \pi G \mu$ \cite{deficit}. This leads to line
discontinuities in CMB anisotropy maps \cite{KS}. The cosmological
fluctuations induced by a network of cosmic strings are ``active''
rather than ``passive'' (as they are for example in the case
of inflationary cosmology). Thus, in spite of the fact that
the power spectrum of fluctuations induced by strings is
scale-invariant (see e.g. \cite{BT}), there are no characteristic
oscillations in the angular power spectrum of CMB anisotropies.
The fact that such oscillations have been observed with high
significance allows us to put the constraint on the cosmic
string tension mentioned in (\ref{constraint}). On the other
hand, recent work \cite{Danos} has shown that limits which are
up to one order of magnitude stronger may be found by analyzing
CMB data directly in position space making use of edge detection
algorithms \cite{Amsel}. 

Long string segments moving through the primodial dark matter
and baryonic gas lead to wedge-shaped overdensities in their
wake. These are called ``wakes'' \cite{wakes}. A string at
time $t$ will produce a wake which is
extended (typical length scale $t$) in the plane of the string world 
sheet, and has a mean thickness of $4 \pi G \mu t$. The initial
density in the wake is twice the background density. The
wakes will grow in thickness by gravitational accretion, and
will lead to signatures in CMB polarization \cite{Holder}
and in 21cm redshift maps \cite{us} (see also \cite{previous}). 
 
The signature of cosmic string wakes in 21cm redshift maps is
a consequence of the overdensity of baryons inside the wake.
This leads to extra emission or absorption of the 21cm radiation
from the direction of the wake \footnote{It is an emission signal
if the gas temperature $T_K$ in the wake is higher than the
temperature $T_{\gamma}$ of the CMB, and is an absorption signal
if $T_K < T_{\gamma}$.}. The signal of a specific string wake produced
at time $t_i$ is seen at a redshift $z(t)$ corresponding to the
time $t$ when our past light cone intersects the wake. It is extended in 
both angular directions (with a length
scale corresponding to the comoving size of the horizon at time $t_i$),
and it is thin in redshift direction direction, with the characteristic
thickness being \cite{us}
\be
\frac{\delta \nu}{\nu}  \, \sim \,  \frac{H w}{c} \, ,
\ee
where $H$ is the expansion rate of space and $w$ is the wake width
(taking into account its increase by gravitational clustering
between the time $t_i$ and $t$), 
both evaluated at the redshift $z$ when the past light cone
intersects the wake. 

21cm surveys provide a window to explore the dark ages of the
universe since they are sensitive to the distribution of
baryons in the universe rather than stars. Most of the baryonic
matter between the time of recombination and that of reionization
is in the form of neutral hydrogen. Neutral hydrogen has a
21cm hyperfine transition line. If we look back at the radiation
from the direction of a baryon cloud, we will see extra emission
or absorption in this 21cm line. If the gas temperature $T_K$ is
higher than that of the microwave photons at the redshift $z$
when our past light cone intersects the cloud, the signal
will be in emission, if $T_K < T_{\gamma}$ it will be in
absorption \footnote{For a detailed review article on the cosmology
with the 21cm line see \cite{Furlanetto}.}. 

A significant advantage of cosmology with the 21cm line compared
to CMB cosmology is that 21cm surveys provide three dimensional
maps of structure in the universe. We map out in two directions
in the sky plus in redshift direction. The effects of gas clouds
at different redshifts which our past light cone intersects will
be seen at different frequencies, namely the 21cm frequency 
shifted by the respective redshifts of intersection. Applied to
cosmic string models, whereas the effects of different strings
which our past light cone intersects are all projected into
the same two-dimensional maps, the 21cm signals will remain well
separated if the strings are at different redshifts.

In previous work \cite{us} the 21cm signal of an individual
string wake was studied, and in a followup paper \cite{spectrum}
the two dimensional power spectrum at fixed redshift was
computed for a scaling distribution of string wakes. In this
paper, we will study the signature of a cosmic string loop.

In the case of cosmic string wakes, we \cite{us} found that
the signal in 21cm redshift surveys is a wedge in the three-dimensional
redshift maps which is extended in the two angular directions,
the angular extent given by the comoving scale of the Hubble
radius at the time $t_i$ that the string wake was formed,
and narrow in redshift direction, the relative thickness being
proportional to $G \mu$ multiplied by the linear growth factor
$z(t_i) / z(t)$ between the redshift $z(t_i)$ corresponding to
the formation time and the redshift $z(t)$ at which the wake
is being seen (at which our past light cone intersects it).
For redshifts $z(t)$ larger than that of reionization and for
values of $G \mu$ comparable to or smaller than the current
observational upper bound (\ref{constraint}), the 21cm signal
is an absorption signal with a brightness temperature
amplitude of up to $400 {\rm{mK}}$, the amplitude
being to a first approximation independent of $G \mu$.

Here, we find that the amplitude of the cosmic string
loop signal can be even larger than the amplitude
for a string wake. However, both the amplitude and
the size of the string-induced feature in the 21cm sky
depends on $G \mu$. Since the accretion pattern
induced by a string loop is roughly spherical, the
induced 21cm signal will be an ellipsoidal region
of extra 21cm emission. Thus, even though the amplitude
of the signal may be larger than that of a cosmic string
wake, it will be harder to observationally distinguish
from foreground noise.

\section{21cm Signal of a Cosmic String Loop}

\subsection{Accretion of Matter onto a Cosmic String Loop}

We first study the accretion of matter onto a cosmic string
loop. We consider a string loop which was produced at time
$t_i \geq t_{eq}$ (and corresponding redshift $z_i$).
As a first approximation, we will replace the string
loop of length $l$ by a point source of mass $m_s = l \mu$ fixed
at a point in space (which we take to be the center of the
coordinates we use to study the accretion process). We then
consider the motion of mass shells surrounding the point mass
shell in response to the gravitational shell, making use
of the Zel'dovich approximation \cite{Zel}. 

As just mentioned, we model the gravitational effects
of the cosmic string loop in terms of a point mass with
associated energy density
\begin{equation}
\rho_{string} = m_s \delta(\bf{x}) \, .
\end{equation}
We will assume radial symmetry. In terms of the radial coordinate
$h$ the above energy density source becomes
\begin{equation}
\rho_{string} = m_s \frac{1}{4 \pi h^2} \delta(h) \, . 
\end{equation}

Let us now consider a mass shell surrounding the string
loop. The physical distance of the shell from the mass point is
\be \label{ansatz}
h(q, t) \, = \, a(t) [q - \psi(q, t)] \, ,
\ee
where $q$ is the comoving scale corresponding to the shell
under consideration and $\psi(q, t)$ is the comoving
displacement of this shell acquired in response to the
gravitational force. The cosmological scale factor $a(t)$
is normalized to be one today: $a(t_0) = 1$, where $t_0$
stands for the present time.

In the Zel'dovich approximation the dynamics of this
mass shell is described by Newtonian gravity, i.e. by
\be \label{Zel1}
\ddot{h} \, = \, - \frac{\partial \Phi}{\partial h} \, ,
\ee
where $\Phi$ is the Newtonian gravitational potential
which in turn is determined by the Poisson equation
\begin{equation} \label{Zel2}
\nabla^2 \phi = 4 \pi G (\rho_{bg} + \rho_{string}) \, ,
\end{equation}
where $\rho_{bg}$ is the background energy density.

Inserting the ansatz (\ref{ansatz}) into the basic
equations (\ref{Zel1}) and (\ref{Zel2}) of the Zel'dovich
approximation and linearizing in $\psi$ yields the
following equation of motion for the comoving displacement
$\psi(q, t)$
\begin{equation}
\ddot{\psi} +\frac{4}{3} t^{-1} \dot{\psi} - \frac{2}{3} t^{-2} \psi 
=  \frac{m_s G t_0^2}{ q^2 t^2} \, ,
\end{equation}
with initial conditions that both $\psi$ and ${\dot{\psi}}$ vanish
at the time $t_i$ when the string loop is formed.
This equation can be solved by means of a Born approximation.
Since this is a standard calculation (see e.g. 
\cite{VilShell, Periv}) we only quote the final result:
\begin{equation}
\psi = - \frac{3}{2} \frac{m_s G t_0^2}{q^2} 
[ 1 - \frac{3}{5} ( \frac{t}{t_1})^{2/3}  - \frac{2}{5} (\frac{t}{t_1})^{-1} ] .
\end{equation}

The Zel'dovich approximation is good until the shells ``turn around'', i.e.
until ${\dot{h(q, t)}} = 0$. This condition determines the comoving scale
$q_{nl}(t)$ which is turning around at the time $t$. A simple calculation
yields
\begin{equation} \label{nonlinear}
q_{nl} = ( \frac{9}{5} m_s G t_0^2 )^{1/3} ( \frac{t}{t_1})^{2/9} 
\equiv q_0 \bigl( \frac{t}{t_i} \bigr)^{2/9} \, ,
\end{equation}
where we have defined a quantity $q_0$ which we will use in the following.
Note that this growth as a power of $t^{2/9}$ agrees with the result of
linear perturbation theory that the accreted mass (which is proportional
to $q_{nl}^3$) scales as $t^{2/3}$.

After turnaround, the shell will collapse and virialize at a physical radius
$R_{max}$ which is one quarter of the radius which the shell would have at the
time of turnaround in the absence of gravitational accretion. Thus, the
physical radius of the nonlinear structure seeded by the string loop at time
$t$ is
\be \label{Rmax}
R_{max}(t) \, = \, \frac{1}{4} a(t) q_{nl}(t) \, 
= \, \frac{1}{4} z(t)^{-1} q_0 \bigl( \frac{z_i}{z(t)} \bigr)^{1/3} \, 
\ee
and the total mass which is gravitationally bound is
\be
M(t) \, = \, \frac{4 \pi}{3} q_{nl}(t)^3 \rho_0 \, ,
\ee
where $\rho_0$ is the background energy density at the present
time $t_0$. Inserting (\ref{nonlinear}) and making use of the Friedmann
equation to substitute for $\rho_0$ yields
\be
M(t) \, = \, \frac{2}{5} m_s \bigl( \frac{t}{t_i} \bigr)^{2/3} \, ,
\ee
in agreement with the result expected from linear perturbation theory.

The mean overdensity (in both baryons and dark matter)
in the nonlinear structure induced by a string loop is
\be \label{over}
\rho_{av} \, = \, 64 \rho_0 
\ee
since the radius has shrunk by a factor of $4$ compared to what it would
be for unperturbed Hubble flow. In fact, towards the center of the structure
the relative overdensity in dark matter is higher than at the edges, since the 
shell which virializes at a distance $R < R_{max}$ from the center has
virialized earlier when the background density was larger. The way we
can compute the cold dark matter distribution is to fix $R < R_{max}$,
determine the value of $q$ for the shell which collapses to radius $R$,
the time when that shell has collapsed, and from those data the mass 
$M(R)$ within that shell. The result is $M(R) \sim R^{3/4}$ which leads
to a density scaling 
\be
\rho_{DM}(R) \, \sim \, R^{-9/4} \, .
\ee
The details of this calculation are presented in \cite{Pagano}. They are
the same which led to the conclusion that the velocity rotation
curve about a cold dark matter halo is not flat - it is more peaked at
the center - and that thus a pure cosmic sting model of structure
formation (which is ruled out because such a model does not
yield the observed oscillations in the angular power spectrum of
CMB anisotropies \cite{CSnogo}) would require hot dark matter
\cite{CSHDM}. We are interested, however, in the density of baryons.
Baryons will shock heat and their density distribution is expected
to be uniform, with a relative overdensity given by (\ref{over}).

We close this subsection by returning to the initial assumption we
made, namely taking the entire loop mass to be concentrated
at the origin. A necessary condition for this approximation to be
justified is that the loop radius is smaller than the physical
height of the outmost virialized shell. Replacing the loop radius
by the loop length $l$ yields a slightly more conservative condition
\be
R_{max}(t) \, \gg \, l \, .
\ee
For a loop formed at time $t_i$ whose induced nonlinear structure
is being considered at time $t \gg t_i$ this condition becomes
\be \label{cond}
\bigl( \frac{l}{t_i} \bigr)^2 < \frac{1}{64} ( \frac{9}{5} G \mu ) \bigl( \frac{z_i}{z} \bigr)^4 
\ee
and is hence easier to satisfy for small loops than large ones. Based on
the scaling solution, these are the more numerous ones.
Inserting numbers, if we are interested in considering string loops formed at $t_{eq}$
(corresponding to a redshift $z_{eq} \simeq 3 \times 10^3$)
which have the longest time to undergo gravitational accretion, and redshift $z \simeq 30$,
then the condition (\ref{cond}) is satisfied for strings with tension $G \mu = 10^{-7}$
for all $l \ll t_i$.

\subsection{Gas Temperature in the Nonlinear Structure}

We assume that the temperature of the gas inside the nonlinear structure induced
by the cosmic string loop is characterized by the molecular kinetic energy which the
particles obtain during the infall process. We use the virialization prescription
that the infalling shell ends up with a height which is half of the height at
turn-around, which is also one quarter of the height the shell would have in the
absence of any gravitational accretion. If we denote the velocity of particles just
before the shock as $v_{shock}$, then the gas temperature is given by 
\be \label{temp}
3/2 k_B T = 1/2 m v_{shock}^2 \, ,
\ee
where $k_B$ is the Boltzmann constant which we will usually set to one in the
natural units we are using. Thus, we must now compute $v_{shock}$.

Based on the shell dynamics studied in the previous subsection, the physical
velocity of the shall labelled by $q$ at late times $t$ (times when we can
neglect the decaying mode of $\psi(t)$) is
\begin{equation}
\dot{h} = -\frac{3}{2} \frac{m_s G t_0^2}{q^2} [ \frac{4}{5} \frac{1}{(t_0 t_1)^{2/3}} t^{1/3}] .
\end{equation}
The shock time $t_{shock}$ is given by
\begin{equation}
h(q , t_{shock}) = \frac{1}{2} h_{max} ( q, t_{nl}) \, ,
\end{equation}
where $h_{max}$is $h_{max} = 1/2 a(t_{nl}) q$. We therefore have
\begin{equation}
h( q , t_{shock}) = \frac{1}{4} a(t_{nl}) q \, .
\end{equation}

Based on our expressions for $h(q, t)$ it is straightforward to compute the time $t_s$
when the shell which turns around at time $t_{nl}(q)$ hits the shock. The result is
that to a good approximation
 \begin{equation}
t_s^{2/3} = t_{nl}^{2/3}( \frac{1 + \sqrt{2}}{\sqrt{2}})  \, .
\end{equation}
Hence, the velocity at the shock is given by
\bea
\dot{h} &=&  -{ (\frac{1 + \sqrt{2}}{\sqrt{2}}})^{1/2} \frac{6}{5} ( \frac{9}{5})^{-2/3} (m_s G t_0^2)^{1/3} 
\nonumber \\
& & \,\,\, \times \frac{ t_1^{4/9}}{( t_0 t_1)^{2/3}} t^{-1/9} \, .
\eea
Making use of (\ref{temp}), this leads to a gas temperature of
\bea
T_K(t) &=&   (\frac{1 + \sqrt{2}}{\sqrt{2}}) m_{HI} \frac{12}{25} ( \frac{9}{5})^{-4/3} (m_s G t_0^2)^{2/3} 
\nonumber \\
& & \,\,\, \times \frac{ t_1^{8/9}}{( t_0 t_1)^{4/3}} t^{-2/9} \, ,
\eea
or, multiplying out the factors of order unity,
\begin{equation} \label{temp1}
T_K(t) =  [0.4] m_{HI}  (m_s G)^{2/3} t_1^{-4/9} t^{-2/9} \, .
\end{equation}

The gas temperature $T_K$ needs to be compared with the temperature $T_{\gamma}$
of the background CMB photons to determine whether the 21cm signal from the
string loop is in emission or in absorption. As a first step, we must insert the
string loop mass $m_s$ into (\ref{temp1}). We express the string loop mass as a
function of the ratio of the string length $l$ to the formation time $t_i$:
\be
m_s G \, = \, \frac{l}{t_i} G \mu t_i \, ,
\ee
and thus
\be
T_K(t) = 0.4 \times m_{HI} (G \mu)^{2/3} \bigl( \frac{l}{t_i} \bigr)^{2/3} z(t)^{1/3} z_i^{-1/3} \, .
\ee
The background photon temperature is given by
\be
T_{\gamma}(t) \simeq 3 \times 10^{-13} {\rm{GeV}} z(t) \, ,
\ee
and hence the ratio of temperatures is
\bea
\frac{T_K(t)}{T_{\gamma}(t)} &\simeq& \frac{m_{HI}}{3 \times 10^{-13} {\rm{GeV}}}
 (G \mu)^{2/3} \bigl( \frac{l}{t_i} \bigr)^{2/3} z(t)^{-2/3} z_i^{-1/3} \nonumber \\
&\simeq& 3 \times 10^8  
(G \mu)_6^{2/3} \bigl( \frac{l}{t_i} \bigr)^{2/3} z(t)^{-2/3} z_i^{-1/3} 
\eea
(where $(G \mu)_6$ indicates the value of $G  \mu$ in units of $10^{-6}$).
Evaluating this result for string loops formed at $z_i = 10^3$ and observed at
$z^{2/3} = 10$ we obtain
\be \label{temp2}
\frac{T_K(t)}{T_{\gamma}(t)} \simeq 3 \times 10^6 \alpha^{2/3} (G \mu)_6^{2/3}
\ee
for the largest loops present at time $t_i$ and
\be \label{temp3}
\frac{T_K(t)}{T_{\gamma}(t)} \simeq 3 \times 10^2 \gamma^{2/3} (G \mu)_6^{4/3} 
\ee
for loops which dominate the distribution at the time $t_i$ which have radius
$R = \gamma G \mu t_i$ at that time \footnote{Loops with radius $R < \alpha t_i$
at time $t_i$ were formed earlier and so started accreting earlier if $t_i > t_{eq}$. For such
loops (\ref{temp3}) is not the correct formula to use. Instead, one should use the
analog of (\ref{temp2}) for the earlier initial time. However, since we are taking $z_i \sim z_{eq}$.
and small loops present at that time only start the gravitational accretion process at the time
of equal matter and radiation, then (\ref{temp3}) is the equation which is
applicable.}

The main message from this subsection is that the gas temperature in the
nonlinear structure formed by a string loop is typically larger than that of the
CMB photons. Hence, the 21cm signal is an emission signal. This is a difference
compared to the string wake case \cite{us}, where for interesting values of
$G \mu$ the 21cm signal will be in absorption. The difference is due to the
fact that string loops exert a stronger gravitational attractive force than that
induced by string wakes.

\subsection{Brightness Temperature}

Consider the CMB radiation at frequency $\nu$ in direction of a gas cloud.
The brightness temperature in this direction due to 21cm emission for a frequency 
$\nu$ is denoted by $T_b(\nu)$ and is given by (see \cite{Furlanetto})
\begin{equation}
T_b(\nu) = T_S(1 - e^{-\tau_\nu})+ T_\gamma(\nu) e^{-\tau \nu}
\end{equation}
where $T_\gamma$ is the background CMB temperature, $\tau_\nu$ is the 
optical depth and $T_S$ the spin temperature of the gas. The first term describes 
the extra emission as a consequence of the hot gas, the second term represents 
the absorption of the background CMB radiation by the gas cloud.

Because UV scattering is insignificant in our case, the spin temperature $T_S$ 
depends only on the gas temperature $T_K$ of the wake and the collision coefficient 
$x_c$:
\begin{equation}
T_S = \frac{1 + x_c}{1 + x_c\frac{T_\gamma}{T_K}}T_\gamma \, .
\end{equation}
The collision coefficient $x_c$ describes the rate at which hydrogen atoms and electrons 
are scattered. We will use values of $x_c$ for which are of observational interest. 
The spin temperature $T_S$ itself describes the relative number density 
\be
\frac{n_1}{n_0} = 3exp(-T_*/T_S)
\ee
 of atoms in the two hyperfine states that when excited produce the 21cm radiation. 
 The quantity $T_* = 0.068K$ is the temperature that corresponds to the energy splitting 
 between these two states. Furthermore, $n_1$ and $n_0$ are the individual number 
 densities of atoms in the hyperfine states. As before, $\tau_\nu$ is the optical depth 
 which can be determined by computing the absorption coefficient of the light way along 
 the gas cloud. The frequency $\nu$ is the blue shifted frequency at the position of the 
 cloud corresponding  to the observed frequency $\nu_0$. The term proportional to 
 $T_\gamma$ is due to absorption and stimulated emission.

As we have mentioned, what is of interest to us is the comparison of the temperature
observed today in direction of the hydrogen gas cloud of the defect as compared to the 
background temperature. This is given by
\begin{equation}
\delta T_b(\nu) = \frac{T_b(\nu) - T_\gamma (\nu)}{1 +z} \simeq \frac{T_S - T_\gamma}{1 + z}\tau_\nu \, ,
\end{equation}
where we have assumed in the last step that the optical depth is smaller than 1 and that one
thus Taylor expand the exponential factor to linear order in $\tau_\nu$
\footnote{However in practice this result is skewed by the intergalactic medium which can 
also be considered a hydrogen gas cloud.}. The factor $1 + z$ comes from the redshifting
of temperatures between the time of emission and the present time.
We can now express $T_{\gamma}$ in terms of $T_S$  to obtain
\begin{equation}
\delta T_b(\nu) \simeq T_S \frac{x_c}{1 + x_c}\frac{\tau_\nu}{1 + z}(1 - T_\gamma/T_K) .
\end{equation}

The optical depth for a general cloud of hydrogen is (in natural units) \cite{Furlanetto}
\begin{equation}
\tau_\nu = \frac{3 A_{10}}{4 \nu^2} \frac{N_{HI}}{4}\frac{\nu_{10}}{T_S} \phi (\nu) \, ,
\end{equation}
where $A_{10}$ is the spontaneous emission coefficient of the
21cm transition, $N_{HI}$ is the column density of hydrogen within the
gas cloud which the photons reaching us pass through, and $\phi (\nu)$ is the line
profile. The hydrogen column density is given by the number density $n_{HI}$ of
hydrogen atoms in the cloud multiplied by the physical diameter $2 R$ of the
cloud, where $R$ is the height computed in the first subsection \footnote{Note that
$n_{HI}$ is the hydrogen number density at the time $t$.}:
\begin{equation}
N_{HI} = 2 n_{HI}^{string} R \, .
\end{equation}
The line profile describes the broadening of the emission lines as a consequence
of the spatial extent of the gas cloud and the resulting redshift difference 
$\delta \nu$ in the frequency
\begin{equation}
\frac{\delta \nu}{\nu} = 2 H R \, ,
\end{equation}
where $H$ is the Hubble expansion constant. Because $\phi (\nu)$ is normalized 
to unity we find that
\begin{equation}
\phi (\nu) = \frac{1}{\delta \nu} 
\end{equation}
for $\nu$ $\epsilon$ $[\nu_{10} - \delta \nu /2$, $ \nu_{10} + \delta \nu /2]$ 
and $\phi(\nu) = 0$ otherwise. This frequency shift plays an integral role in the
observability of the 21cm signal since it determines the width of the string signal 
in the redshift direction.

Note that the dependence on $G \mu$ cancels out between the column height and
the line profile, as it did in the case of cosmic string wakes \cite{us}. Thus, we
obtain
\begin{equation}
\tau_\nu = 3 \frac{A_{10}}{16 \nu^2} \frac{1}{T_S}
\frac{n_{HI}}{H_0 \Omega_m^{1/2} ( 1 + z)^{3/2}} \, ,
\end{equation}
where we have used $H(z) = H_0 \Omega_m^{1/2} (1 + z)^{3/2}$ in which 
$\Omega_m$ is the ratio between the matter energy density and the
critical density (the density for a spatially flat universe). We can insert this 
into the expression for $\delta T_b ( \nu)$ to obtain the signal strength
\begin{equation}
\delta T_b(\nu) = \frac{x_c}{1 + x_c}(1 - \frac{T_\gamma}{T_K}) 
\frac{3 A_{10}}{16 \nu^2}\frac{1}{T_s} \frac{n_{HI}^{string}}{ H_0 \Omega_m^{1/2}}( 1 +z)^{-5/2} \, .
\end{equation}
We can remove the dependence on the hydrogen number density $n_{HI}$ by 
introducing the background density $n_{bg}$ and noting that the overdensity 
of hydrogen $n_{HI}^{string}$ is $64$ times that of the background. This
yields
\begin{equation} \label{btemp1}
\delta T_b(\nu) = \frac{x_c}{1 + x_c} (1 - \frac{T_\gamma}{T_K}) \frac{3}{16 \nu^2} A_{10} \frac{n_{HI}^{string}}{n_{bg}} \frac{n_{0, bg}}{H_0 \Omega_m^{1/2}}( 1 +z)^{1/2} \, ,
\end{equation}
where we have rescaled the background number density to the present time. 

In the expression (\ref{btemp1}) for the brightness temperature, the string tension
$G \mu$ enters only in two minor ways. Firstly, it determines the gas temperature
$T_K$, but since $T_K \gg T_\gamma$ for the parameter values we are interested
in, this dependence is negligible. The second place where $G \mu$ enters is
in the collision coefficient $x_c$ which depends on the gas temperature and
hence on $G \mu$. However, for our parameter values $x_c \gg 1$ and hence
this second dependence is negligible as well.

Inserting the values  $H_0 = 73 km s^{-1} Mpc^{-1}$, 
$A_{10} = 2.85 \times 10^{-15} s^{-1}$, $v_{10} = 1420 MHz$, 
$\Omega_m = 0.26$, $T_* = 0.068 K$, and
$n_{bg} = 1.9 \times 10^{-7}cm^{-3} ( 1 + z)^3$ we obtain
\begin{equation}
\delta T_b ( \nu) \simeq [1.1] \frac{x_c}{1 + x_c} ( 1 + z)^{1/2}
\end{equation}
where the collision coefficient $x_c$ is given by
\begin{equation}
x_c = \frac{n \kappa^{HH}_{10} T_*}{A_{10} T_\gamma}
\end{equation}
and $\kappa^{HH}_{10}$ is the de-excitation cross section whose value is 
$2.7 \times 10^{-9} m^3s^{-1}$ at high temperatures corresponding to 
$G \mu \sim 0.3 \times 10^{-6}$ at a redshift of $( 1 + z) \sim 30$ and 
$( 1 + z_1) \sim 10^3$. These values lead to the collision coefficent of 
$x_c \gg 1$.  We have also made use of the fact that the 
overdensity of the hydrogen gas is $64$ times that of the background. 
The ensuing signal of the cosmic string loop is then 
\begin{equation}
\delta T_b ( \nu) \sim 6 K \, ,
\end{equation}
which is an emission signal in the 21cm map. This temperature is even
larger than that obtained for cosmic string wakes, the reason being that
the overdensity inside of a string loop-induced structure is larger than
that in a string wake by a factor of $16$.

\subsection{Geometry of the Signal}

Projected onto the two angular directions, the 21cm signal of a cosmic
string loop looks like a filled circlular region of extra 21cm emission.
The angular scale $\theta$ of this region is determined by the ratio of the 
comoving distance $\ell_{com}$ corresponding to $R_{max}$ and the 
distance of the observer's past light cone. We can express this as
\begin{equation}
\frac{\theta}{180^\circ} = \frac{\ell_{com}}{t_0} \, .
\end{equation}
The comoving length is related to the physical distance by $a(t)$: 
\be
\ell_{com} = a^{-1}\ell_{phys} = ( 1 + z) R_{max} \, , 
\ee
and thus the angular size of the defect is determined by the maximum radius 
of the nonlinear region which has collapsed onto the cosmic string loop. 
Thus,
\begin{equation}
\frac{\theta}{180^\circ} = \frac{ ( 1 + z) R_{max}(z)}{t_0} \, .
\end{equation}

Starting from (\ref{Rmax}) and inserting the mass of the string loop in terms of the
loop length $l_s$ we obtain
\be \label{Rmax2}
R_{max}(z) = \frac{1}{4} (\frac{9}{5})^{1/3}(G \mu)^{1/3} \bigl( \frac{l_s}{t_i} \bigr)^{1/3}
t_0 z_i^{-1/6} z^{-4/3}
\ee
The angular size therefore becomes 
\begin{equation}
\theta \simeq  180^{\circ} \times \frac{1}{4} (\frac{9}{5})^{1/3}( G \mu)^{1/3}  
\bigl( \frac{l_s}{t_i} \bigr)^{1/3} z_i^{1/6} z^{-1/3} \, .
\end{equation}
Inserting $l_s = \alpha t_i$ (valid for the largest loops
present at time $t_i$), $z_i = 10^3$ and $z = 30$ we obtain
\begin{equation}
\theta \sim 2^\circ \times 10^{-1} (G \mu)_6^{1/3} \alpha^{1/3} \, .
\end{equation}
Smaller loops at $t_i$ give a correspondingly smaller angular scale.

The redshift extent of the blob of extra 21cm emission coming from the
nonlinear structure seeded by a string loop was derived previously. It is
\begin{equation}
\frac{\delta \nu}{\nu} = 2 H R_{max}(z) \, .
\end{equation}
Inserting (\ref{Rmax2}) and using the same parameters as above we obtain
\be
\frac{\delta \nu}{\nu} \sim 2 \times 10^{-3} (G \mu)_6^{1/3} \alpha^{1/3} \, ,
\ee
which is a larger value than the thickness for the 21cm signal of
a cosmic string wake. The reason for the larger value is that in the
case of a string wake (planar accretion) the height of the nonlinear
structure scales linearly in $G \mu$ whereas for a string loop (spherical
accretion) it scales as $(G \mu)^{1/3}$.

\section{Discussion}

In this paper we have discussed the 21cm signature of cosmic strings loops.
We found that for values of the string tension close to the current upper
bounds, the signature is an ellipsoidal region of extra 21cm emission.
The brightness temperature of this signal is several degrees K for string
loops produced at a redshift of $z_i = 10^{3}$ and observed at a
redshift of $z = 30$. The brightness temperature is to a first approximation
independent of the string tension, as in the case of the signature of a
cosmic string wake.

Projected onto the celestial sphere, the angular extent of the region
of extra 21cm emission is (for the same redshifts $z_i$ and $z$
mentioned in the previous paragraph) $0.2^{\circ}$ for a value of 
$G \mu = 10^{-6}$ and for a string loop with length $l_s = t_i$.
This angle scale as $(G \mu)^{1/3}$ and also as $l_s^{1/3}$. In
redshift direction the relative thickness of the region of extra 21m emission
is $\frac{\delta \nu}{\nu} \sim 2 \times 10^{-3}$ (for the same values
of $G \mu$ and $\alpha$ as above), and this thickness has the same
scaling in both $G \mu$ and $l_s$ as the angular scale.

In terms of amplitude and both angular and redshift extent the
cosmic string signal should be easily detectable by the SKA 
experiment, and possibly even with LOFAR. However, the
signal might be hard to disentangle from noise and foregrounds,
in contrast to the signal of a string wake which forms a very specific
geometrical pattern.

A scaling distribution of string loops will lead to a superposition of
many ellipsoidal regions of extra 21cm emission. However, since 21cm redshift
surveys will produce three dimensional maps, the signatures of
individual string loops will be spaced further apart as the corresponding
signals in CMB anisotropy maps would be. Hence, we do not expect
the string loop signal to become Gaussian as a consequence of the
central limit theorem. 

Since the string loop distribution can be viewed as statistically
independent from the distribution of string wakes, the 21cm signature
of a scaling string model will be the linear superposition of what we
found in the current paper with what was found in \cite{us}. It
would be interesting to compute the total angular power spectrum
of the 21cm signal, following what was done in \cite{spectrum}.
However, in computing a two dimensional correlation function, and in
doing this for a projected map, a lot of discriminatory power is lost.
An improvement could be obtained by computing a three-dimensional
power spectrum. Even more promising, however, would be to find
position space algorithms to search for string signatures.

Our analysis is applicable with very few changes to a second type
of topological defect, namely to global monopoles, pointlike
defects arising in a theory in which a global symmetry is broken
in a way that the second homotopy group of the vacuum manifold
is non-trivial \cite{monopole}. 
Note, however, that theories with broken global 
symmetries have potential problems and are hence not
considered as interesting as those with broken local
symmetries \cite{problems}. Global monopoles also give
rise to a spherical accretion pattern and to nonlinear density
perturbations at early times. The density distribution of a monopole
is different from that of a cosmic string loop since field gradient energy
extends arbitrarily far from the center of the monopole \footnote{The total
energy is in fact logarithmically divergent as a function of the distance from
the center. This is not a concern in the context of cosmology since
there is an effective infrared cutoff at the Hubble scale and the energy
inside the Hubble radius remains a fraction $G \eta^2$ of the background
energy.}. As a consequence, the dark matter distribution has a
different radial profile compared to that of the structure induced
by a cosmic string loop. However, the baryon distribution inside the
region which has undergone shock heating will be identical for string
loops and global monopoles, and hence the 21cm signal of a 
single global monopole will not be distinguishable from that of
a cosmic string loop (see \cite{Pagano} for details).
On the other hand, the scaling distribution of
monopoles will differ from that of string loops, and hence the overall
21cm maps can in principle be distinguished.

\begin{acknowledgments} 
 
This work is supported in part by a NSERC Discovery Grant and by funds from the 
CRC Program.

\end{acknowledgments}

\end{document}